\documentclass[12pt,preprint]{aastex}
\def\HI {H\kern0.1em{\sc i}} 
\def\radm {rad m$^{-2}$} 
\def\etal   {{\sl et~al.\ }}
\def\deg{$^{\circ}$}

\begin{document}

\title{~~\\ {VLBA Polarimetry of Three Powerful \\ Radio Galaxy Cores}}
\author{G. B. Taylor\altaffilmark{1}, D. H. Hough\altaffilmark{2} 
\& T. Venturi\altaffilmark{3}}

\altaffiltext{1}{National Radio Astronomy Observatory, Socorro, NM
  87801, USA}
\email{gtaylor@nrao.edu}
\altaffiltext{2}{Trinity University, San Antonio, TX 78212, USA}
\email{dhough@trinity.edu}
\altaffiltext{3}{Istituto di Radioastronomia, CNR, Bologna, Italy}
\email{tventuri@ira.bo.cnr.it}




\shorttitle{VLBA Polarimetry of Radio Galaxies}
\slugcomment{Accepted to the Astrophysical Journal}

\begin{abstract}

We present sensitive, high dynamic range, Very Long Baseline Array
(VLBA) polarimetric observations of the cores of three powerful radio
galaxies: 3C\,166, 3C\,236 and 3C\,390.3.  Significant polarization is
detected in one source (3C\,166) allowing us to map out the Faraday
Rotation Measure (RM) distribution and projected magnetic field
direction.  The inner jet of 3C\,166 is found to have a rest frame
RM of $-$2300 \radm, similar to those found in quasars cores.  No
polarized flux was detected from the other two sources, but in both
counterjets are seen.  The counterjet in 3C\,236 was previously known,
but the detection in 3C\,390.3 is a new discovery.  We suggest that
the low fractional polarization in radio galaxy cores is the result of
Faraday depolarization by ionized gas associated with the accretion
disk.  The lower polarization of radio galaxy cores compared to
quasars is then naturally explained by unified models as a result of
the viewing angle.

\end{abstract}


\section{Introduction}

The cores of high power (FRII) radio galaxies typically have quite low
fractional polarization ($<$0.6\%) on arcsecond scales (e.g., Rudnick,
Jones \& Fiedler 1986, Rusk 1988).  This is in contrast to the
properties of quasar cores which are generally more highly polarized
on arcsecond scales (2--5\%).  Rudnick \etal\ (1986) have speculated
that the low polarization of radio galaxy cores could be due to
optical depth effects of self-absorbed emission, highly randomized
internal magnetic fields, or Faraday depolarization by a highly
tangled screen.

Multi-frequency polarimetry with the Very Long Baseline Array (VLBA)
telescope has recently revealed absolute Faraday Rotation Measures
(RMs) in excess of 1000 rad m$^{-2}$ in the central regions of 7 out
of 8 strong quasars studied (e.g., 3C\,273, 3C\,279 -- Taylor 1998,
2000).  Beyond a projected distance of $\sim$20 pc, however, the jets
are found to have $|$RM$|$ $<$ 100 rad m$^{-2}$.  The RM measures the
density weighted magnetic field along the line of sight.  Such sharp
RM gradients cannot be produced by cluster or galactic-scale magnetic
fields, but rather must be the result of magnetic fields organized
over the central 1--100 pc.  Taylor (2000) postulates that the cores
and inner jets of quasars are viewed through ionized gas associated
with the nuclear region.  According to the unified scheme (see review
by Antonucci 1993) linking
quasars to FRII radio galaxies, one expects that the jet components of
FRII radio galaxies within $\sim$100 pc of the center of activity will
be viewed through a deep Faraday screen.  This would explain the low
fractional polarization of radio galaxy cores.  With the high
resolution afforded by VLBI observations it might be possible to
resolve the Faraday screen and recover some polarized flux density.
Furthermore, once the jet components move farther from the nuclear
environment the Faraday depth should drop, and it may be possible to
measure both significant polarization from the jet and still high RMs.

Very little is known about the parsec-scale polarization properties of
classical FRII radio galaxy cores.  To investigate their polarization
properties, and to image their RM distribution in the central parsecs,
we observed three FRII radio galaxies having relatively strong cores
and some integrated fractional polarizations: 3C\,166 (B0642+214),
3C\,236 (B1003+351), and 3C\,390.3 (B1845+797).
All three sources are classical FRII radio galaxies, and 3C\,236, with a
size of 4 Mpc, has the added distinction of being one of the giant
radio galaxies.


We assume H$_0 = 70$ km s$^{-1}$ Mpc$^{-1}$ and q$_0$=0.5 throughout.

\section{Observations and Data Reduction}

The observations, performed on 2000 July 29-30, were carried out at
4.517, 4.547, 4.867 and 4.897 GHz (see Table 1) using the 10 element
VLBA and a single antenna from the VLA\footnote{The National Radio
Astronomy Observatory is operated by Associated Universities, Inc.,
under cooperative agreement with the National Science Foundation.}.
Unfortunately, due to a hardware fault, the VLBA antenna at Los Alamos
was unable to participate in this experiment.
Right- and left-circular polarizations 
were recorded using 2 bit sampling across bandwidths of 4 MHz at each
frequency.  The 
VLBA  correlator produced 16 frequency channels across each 
4 MHz band during every 2.1 s integration.

Amplitude calibration for each antenna was derived from measurements
of the antenna gain and system temperatures during the run.  No
atmospheric opacity corrections were performed.  Fringe fitting was
performed on the strong calibrator 1928+738 to determine the delays to
each station, and on 3C\,279 to determine the R-L delay difference
(caused by differences in the electronic path length between the two 
orthogonally polarized receivers) on
the reference antenna (FD).  Failure to properly set the R-L delay 
would cause the polarization angles to appear to drift with time.
No global fringe fit was performed as the
solutions found were consistent with 0 residual delay to within the
scatter of those solutions.

The data were averaged over 20 s intervals and edited and
self-calibrated using AIPS and DIFMAP (Shepherd, Pearson \&
Taylor 1994, Shepherd 1997) in combination.  Feed polarizations of the
antennas were determined using the unpolarized calibrator
OQ\,208 and the AIPS task LPCAL. We assumed that the VLBA
antennas had good-quality feeds with relatively pure polarizations,
which allowed us to use a linearized model to fit the feed
polarizations.  The absolute polarization angle calibration was
performed by comparing the VLBA measurements of the calibrators
1308+326 and 1749+096 to near simultaneous measurements made with the
VLA as part of the VLA polarization monitoring program
(http://www.aoc.nrao.edu/$\sim$smyers/calibration/).  Both the VLA and
VLBA data (after correction) at 4867 MHz, are given in Table 2.  As a
check, we also imaged 3C\,279 in I, Q and U.  This quasar has a
strongly polarized jet component (C4) with a moderately low Faraday
Rotation Measure (RM).  The polarization angle and RM of this
component change slowly with time (Zavala \& Taylor 2001) and our
measurement of $-$84\deg\ is within a few degrees of other 
observations carried out
at 8 GHz and above in the summer of 2000 (R. Zavala, private
communication).

\section{Results}

We present total intensity images for 3C\,166, 3C\,236 and 3C\,390.3.
Polarized emission was detected only in 3C166, and for this source
we further present its polarization properties and RM distribution.

\subsection{3C\,166}

This radio source was identified by Bridle \& Fomalont (1978) with a
17.7 magnitude galaxy at z=0.2449 in a cluster (Wyndham 1966).
Spangler \& Bridle (1982) showed the radio structure to have a
classical double morphology, but with rather dissimilar radio lobes.
Further high resolution VLA observations confirmed that 3C\,166 has an
FRII morphology with edge brightened lobes extending north-south over
50 arcsec (170 kpc) (Leahy \& Williams 1984; Neff, Roberts \&
Hutchings 1995).  No evidence of jets is seen in any of the VLA
images. The core is quite prominent, containing 475 mJy
(Rudnick \etal\ 1986) at 5 GHz compared to the integrated flux
density of 970 mJy (Gregory \& Condon 1991; Zukowski \etal\ 1999).
We chose to observe 3C\,166, primarily based on the relatively high
fractional polarization of the core of 1.7\% (8 mJy of polarized flux
density) at 5 GHz (Rudnick \etal\ 1986).

On the parsec scale, we recover 395 mJy (83\%) of the kiloparsec-core flux
density at 5 GHz.  We assume the parsec-scale core is the compact
component at the very southern end of the jet.  Model-fitting 
with elliptical Gaussians shows this component
contains 135 mJy, and the rest of the emission is in a one-sided jet
that starts out in position angle $-$26\deg\ and curves gradually
toward due North (Fig.~1).  A total of 4 mJy of linearly polarized
flux is detected on the parsec scale, mostly from the jet.  The core
is very weakly polarized (0.2\%), but the polarization increases
quickly in the inner jet (2\%), and reaches 12\% 8.5 mas out from the
core.  The electric vector polarization angle (EVPA) is shown in
Fig.~1 to be fairly well ordered and transverse to the jet direction.
The same EVPA was found for the VLA core by Rudnick \etal\ (1986).
By fitting the EVPA as a function of wavelength-squared we derive the
RM distribution (Fig.~2).  The RMs in the jet have a mean value of 32
$\pm$ 30, basically consistent with little or no RM contribution from
the host galaxy.  After correcting for the RM, the projected magnetic
field direction beyond $\sim$10 pc from the core is very nearly
parallel to the jet direction.  The inner jet, just north of the core,
however, exhibits substantial RMs of $-$1500 \radm.  Assuming these
RMs are produced in the rest frame of the source then correcting by
(1+z)$^2$ yields an intrinsic RM of $-$2300 \radm.  After correcting
for the RM, the projected magnetic field in the inner jet is nearly
perpendicular to the axis of the jet.  A similar transition from a
perpendicular to a parallel magnetic field configuration, with similar
RM in the inner jet, has been observed in the nucleus of
the lobe-dominated quasar 3C\,245 (Hough, Barth, \& Yu 1998; 
Hough, Homan, \& Wardle 2001).

There is no sign of any counterjet emission to the southeast over a
comparable distance from the core as the jet extends. We can estimate
an upper limit to the counterjet emission over a rectangular region
matched to the outer northwest jet that contains 30 mJy. Using a region
covering 130 beam areas (820 ${\rm mas}^{2}$) and the rms image noise
of 80 $\mu$Jy per beam, we find an upper limit of 10 mJy to the
counterjet flux density.  Thus the lower limit on the
jet-to-counterjet brightness ratio is 3.  Within 3 mas of the core the
jet-to-counterjet limit is 50.

\subsection{3C\,236}

   The ``giant'' radio galaxy 3C\,236 was studied on a vast range of
angular scales by Barthel \etal (1985). Observations of 3C\,236 with
the Westerbork Synthesis Radio Telescope at 1.4 GHz (23$''$x13$''$
beam) revealed the presence of twin outer lobes that give the source
an overall angular size of $\sim$42 arcminutes. Adopting the redshift
$z$ = 0.1005 from Hill, Goodrich, \& DePoy (1996), the projected
linear size of 3C\,236 is then 4.4 Mpc, making it the largest radio
galaxy known.  The northwest lobe has a central ridge with a peak that
may be a ``hot spot'' along PA ${304}^{\circ}$, and the southeast lobe
has a weak terminal ``hot spot'' fed by a segment of emission all
along PA ${122}^{\circ}$.  VLBI maps at 1.6 GHz and 5 GHz (Schilizzi
\etal\ 2001) show structures on scales between 1 mas and 1$''$ on both
sides of the compact component, presumed to be the ``core'', that can
be compared to our new 5 GHz VLBA image.  This inner double does not
connect up with the outer hot spots, leading to the classification of
3C\,236 as one of the ``double-double'' radio sources (Schoenmakers
\etal\ 1999, 2000).  O'Dea \etal\ (2001) present HST observations, and
based on dynamical and spectral aging arguments, suggest that the
central engine powering the radio source turned off for $\sim$ 10$^7$
yr, and has only recently been restarted.

     Our VLBA image at 5 GHz is displayed
in Fig.~3.  While the overall structure is suggestive of an
``S''-pattern, the source is decidedly asymmetric. The innermost
structure is that of a typical ``core-jet'', with a 217-mJy core and a
51-mJy one-sided jet extending 25 mas to the northwest. The jet is not
straight, but rather shows undulations in PA $\sim$ ${292}^{\circ}$,
${297}^{\circ}$, ${290}^{\circ}$, ${300}^{\circ}$, and ${294}^{\circ}$
(the latter two being in the outer 10 mas of the jet where it is
faintest before disappearing from view). There is no sign of any
counterjet emission to the southeast over a comparable distance from
the core. In fact, we can estimate an upper limit to the counterjet
emission over a rectangular region matched to the northwest jet. Using
a region covering 40 beam areas (210 ${\rm mas}^{2}$) and the rms
image noise of 93 $\mu$Jy per beam, we find an upper limit of 3.8 mJy
to the inner counterjet flux density.  Thus the lower limit on the
jet-to-counterjet brightness ratio is 13.5.  The Barthel \etal\ (1985)
5 GHz VLBI map shows the first few milliarcseconds of the jet in the
same orientation as our image (a ``counterjet'' feature on their very
low-dynamic-range image is clearly not real).

     On the northwest side, there is a gap of nearly 40 mas between the 
end of the inner jet and a diffuse 41-mJy component. The center of this
diffuse component (B1) lies about 70 mas from the core (B2) in 
PA $\sim$ ${304}^{\circ}$ and has a mean diameter of $\sim$20 mas. Its 
position angle is thus significantly greater than the mean position 
angle of the inner jet ($\sim$ ${292}^{\circ}$). 

     On the southeast side, there is a 20-mJy bent counterjet segment
that begins about 41 mas from the core in PA $\sim$ ${111}^{\circ}$
and ends about 59 mas away in PA $\sim$ ${118}^{\circ}$. Beyond this,
there is a large extended region at $\sim$130 mas in PA $\sim$
${120}^{\circ}$.  Due to insufficient short baselines we cannot
properly estimate the flux density of this extended component, or be certain
that we have faithfully reproduced its morphology (Fig.~4).  The shape
and location of this component are in rough agreement with a 1.6 GHz
EVN+MERLIN image (Schilizzi \etal\ 2001).

     Our northwest diffuse feature corresponds to feature ``B1'' on
the 1.6 GHz VLBI map of Barthel \etal\ (1985), and our curved
counterjet segment and extended feature 130 mas from the core
correspond to portions of their elongated feature ``C''. The 1.6 GHz
VLBI map also exhibits a diffuse halo in which these central features
are imbedded, and a compact component ``A'' in a large diffuse patch
nearly 1$''$ to the northwest.
The 1.6 GHz images of Schilizzi \etal\ (2001) show the same features that
we see on our 5 GHz VLBA image, with more low-level emission in the
vicinity of our outer 130 mas counterjet side feature and further 
low-level emission 500 mas to the northwest.

      The overall picture in 3C\,236 seems to suggest an initial jet (and a
presumably invisible counterjet) that, while undulating, have an 
average position angle $\sim$ ${10}^{\circ}$ less than emission that 
appears beyond the jet and counterjet gaps. This suggests some 
mechanism on the scale of 25 mas ($\sim$40 pc) is responsible for jet 
bending which redirects the flows in directions that - quite remarkably - 
point directly to large-scale features 2 Mpc away on either side of 
the source. 

      No polarized flux was detected in our 5 GHz VLBA observations.
The polarized intensity image has an rms noise of 45 $\mu$Jy per
beam. At 5 GHz, Rusk (1988) observed 0.9\% polarization corresponding
to 13 mJy of linearly polarized flux density in the nucleus of
3C\,236.  Since Rusk (1988) presented a 15 GHz VLA image that suggests
the polarized arcsecond-scale emission comes mainly from the
northwestern of two components in a close 1$''$ double, we must
conclude that there is very little polarization on the size scales
sampled by our observations. VLBA observations at a lower frequency
may prove useful in detecting this polarized flux on scales
approaching 0.5 to 1.0 arcsecond. It is interesting to note that Rusk
(1988) found the electric vector position angle for the 5 GHz VLA core
of 3C\,236 to be ${291}^{\circ}$, well-aligned with the main
structural axis of the source.

\subsection{3C\,390.3}

3C\,390.3 is a broad line nearby FRII radio galaxy (z=0.0561 -- Hewitt
\& Burbidge 1991), characterized by prominent lobes and hot spots with
fine radio structure (Leahy \& Perley 1995).  Alef \etal\ (1996)
detected a couple of faint jet features between the core and the
northwest lobe on their VLA image; there is also a bright portion of
the jet visible in the lobe feeding a compact hot spot.  This source
was also one of the first sources to be found to exhibit double-peaked
emission lines (Eracleous \& Halpern 1994).  By modeling the
double-peaked emission lines with a disk they derive an inclination
for the disk of 26${^{+4}_{-2}}$ degrees.  Assuming that the radio
jets are perpendicular to the disk then the angle between the
line-of-sight and the jet axis, $\theta$, is also 26$^\circ$.  On the
basis of a four epoch imaging monitoring, Alef \etal\ (1996) proposed
superluminal motion of the order of 0.7 mas yr$^{-1}$ along the
one-sided milliarcsecond radio jet, corresponding to $\beta_{\rm app}$
= 2.7 for our choice of the cosmological parameters.

Our naturally weighted VLBA image, shown in Fig.~5, 
is in good agreement with previous results from the literature.
A taper has also been applied to increase sensitivity to some faint
extended components.
The total flux in our image, dominated by 206 mJy in a compact core 
and 120 mJy in a jet, recovers $\sim$94\% of the
core peak flux density in the VLA image shown in Alef \etal\ (1996). 
The milliarcsecond morphology is dominated by the core-jet
structure, aligned in position angle $-35^{\circ}$, 
and consistent with the large scale orientation projected in the 
plane of the sky. The jet is edge-darkened and its brightness
drops considerably at $\sim$ 10 mas from the peak emission.
A uniformly weighted image convolved with a circular beam with
$\theta_{\rm FWHM}$ = 0.8 mas (not shown here) 
shows that the centrally peaked jet in Fig.~5 is actually the result of 
a sequence of knots of similar brightness, almost perfectly 
aligned in p.a. $-35^{\circ}$. A gap in the parsec-scale jet emission 
at $\sim$ 3.5 mas from the peak brightness is also evident. 

The very high sensitivity of the image presented here allows us to
follow the jet out to $\sim$ 20 mas. A diffuse counterjet feature,
significant at 7$\sigma$, is visible in Fig.~5 at $\sim$ 14.7 mas.
At
this distance from the core the jet-to-counterjet brightness ratio is
2.  The region where the counterjet is undetected covers 10 beam areas
(40 ${\rm mas}^{2}$).  Given the rms image noise of 82 $\mu$Jy per
beam, we find an upper limit of 0.8 mJy to the inner counterjet flux
density.  Thus the lower limit on the inner jet-to-counterjet
brightness ratio is 140.

      No polarized flux was detected in our 5 GHz VLBA observations.
The polarized intensity image has an rms noise of 60 $\mu$Jy per
beam.

\section{Discussion}

\subsection{Constraints from Doppler Boosting}


For a simple relativistic beaming model in which the jet and counterjet
are intrinsically identical, the theoretical jet-to-counterjet
brightness ratio is $J$ = ${[(1 - \beta cos \theta)/(1 + \beta cos
\theta)]}^{-(n+\alpha)}$, where $\beta$ is the speed parameter,
$\theta$ is the angle of the jet axis to our line-of-sight, the index
$n$=2 for a continuous jet, and a spectral index $\alpha$ = 0.5
($S \propto {\nu}^{-\alpha}$) is assumed.  

In 3C\,166 the jet brightness drops smoothly along the parsec-scale 
jet. From our images we derived a jet-to-counterjet ratio
changing from 50 at $\sim$ 3 mas, to 32 at $\sim$ 10 mas, and 
finally reaching 4 at $\sim$ 30 mas. The constraint $\beta cos\theta$
in the proximity of the core is 0.67, leading to a maximum
viewing angle $\theta_{\rm max} = 48^{\circ}$.

In 3C\,236, for the inner bright jet segment, we find $\beta cos \theta$ = 0.48
which, as $\beta$ approaches one, gives a maximum angle to the
line-of-sight ${\theta}_{\rm max}$ = ${61}^{\circ}$. On the basis of its
immense projected linear size, Barthel \etal\ (1985) have argued that
the overall source axis must not lie far from the plane of the
sky. The apparently contradictory requirements on the source
orientation based on observations of the smallest and largest scales
in 3C\,236 can be reconciled by assuming overall jet curvature on the
order $\sim$ ${20}^{\circ}$ or more; such curvature may be plausible
in 3C\,236, given the bends of at least ${10}^{\circ}$ in the plane of
the sky discussed in \S3.2.

Under the same assumptions the high brightness asymmetry of the inner
VLBI jet in 3C\,390.3 leads to $\beta cos\theta$ = 0.76, which implies
$\beta_{\rm min}$ = 0.76 and a maximum viewing angle $\theta_{\rm max}
\sim 41^{\circ}$ for $\beta \rightarrow$ 1, consistent with the broad
line nature of the host galaxy.  This limit for $\theta$ is also in
reasonable agreement with that obtained from the superluminal motion
proposed by Alef \etal\ (1996), i.e. $\theta_{\rm max} \sim
40^{\circ}$ for $\beta_{\rm app} = 2.73$. The corresponding
$\beta_{\rm min}$ is 0.94.  The presence of the counterjet feature at
$\sim 14.7$ mas yields $\beta cos\theta \sim 0.14$. The very good
alignment between the parsec and kiloparsec scale morphology suggests
that jet deceleration may be relevant beyond $\sim 10$ mas from the
core, i.e.\ 12 pc projected on the plane of the sky.

We compared our images of 3C\,390.3 to those presented in Alef \etal\
(1996), in order to confirm their proposed superluminal motion.
It is tempting to associate their component located at 5 mas from the
core (see Fig.~3 in Alef \etal\ 1996) with our jet component located
at 9.3 mas, just before the brightness drops.
However the morphology of the milliarcsecond jet in Fig.~5 suggests that
it is actually the blend of various knots, unresolved at the present
resolution, and this makes any component identification uncertain.
In spite of this difficulty, it is interesting to note that the inner
portion of the jet (i.e., before the sharp the drop in brightness)
has clearly stretched over the last 11.28 years, going from
$\sim$ 6 mas to $\sim 10$ mas. This corresponds to a global expansion
of 0.35 mas yr$^{-1}$, i.e. superluminal motion with apparent tranverse
speed $v_{\rm app} = 1.28$c. This constrains the beaming parameters
to $\beta_{\rm min}=0.79$ for $\theta \sim 37^{\circ}$,
and is in good agreement with the limit
derived on the basis of the jet-to-counterjet asymmetry within 
10 mas of the center of activity.  

We note that other explanations for the
observed one-sided morphologies of these radio galaxies 
are also possible.  These include differences in the
environment from side-to-side, and even free-free absorption 
from the circumnuclear torus as observed in 3C\,84 by 
Walker \etal\ (2000).

\subsection{Implications of the RM Structure in 3C166}

Despite all three sources having comparable jet brightnesses, linearly
polarized flux density was only detected on the parsec scale from
3C\,166.  Another intriguing difference between the sources is that
3C\,236 and 3C\,390.3 have detectable counterjets.  In addition, while
we did not detect any polarized emission in the inner 100 pc of the
3C236 jet that we were able to image, Rusk (1988) found significant
arcsecond-scale polarization in a component roughly 1 kpc from the
core.  One way to explain these differences in light of the unified
schemes is to require that the jet in 3C\,166 is oriented more closely
to the line-of-sight than the jets of the other two sources.  This
results in greater Faraday depths towards 3C\,236 and 3C\,390.3, and
causes the counterjet to appear fainter due to Doppler deboosting in
3C\,166.  Even in 3C\,166 the core itself is depolarized and the inner
jet exhibits RMs of $-$2300 \radm.  This is consistent with the
detection of counterjet components in 3C\,236 and 3C\,390.3.  A minor
problem with this picture is that the jet-to-counterjet asymmetry
close to the core is similar for all three sources (see Table 3).
However, if the jet starts out highly relativistic and then slows
after traveling $\sim$20 pc then a counterjet could be detected for
those systems oriented closer to the plane of the sky.  This
possibility is supported by our argument for deceleration in the jets
of 3C\,390.3 (see \S 4.1 above).


\section{Conclusions}

We have detected linearly polarized flux on the parsec scale for one
(3C\,166) of three powerful radio galaxy cores observed.  In 3C\,166
we find substantial RMs ($-$2300 \radm) and RM gradients.  We suggest
that Faraday depolarization is the most likely explanation for the low
fractional polarization of the cores of radio galaxies.  These
conclusions are rather tentative considering the small number of
sources involved in our study.  A better test of this explanation and
unified schemes would be to consider a complete sample of powerful
radio galaxies and quasars chosen on the basis of an unbeamed
property.  One such sample that could warrant further study
are the FRII lobe-dominated quasars from 
the 3CR catalog (Hough, Barth \& Yu 1998, Hough \etal\ 2001).

\acknowledgments

This research has made use of the NASA/IPAC Extragalactic Database (NED)
which is operated by the Jet Propulsion Laboratory, Caltech, under
contract with NASA.

\clearpage


\begin{figure}
\vspace{19cm}
\includegraphics{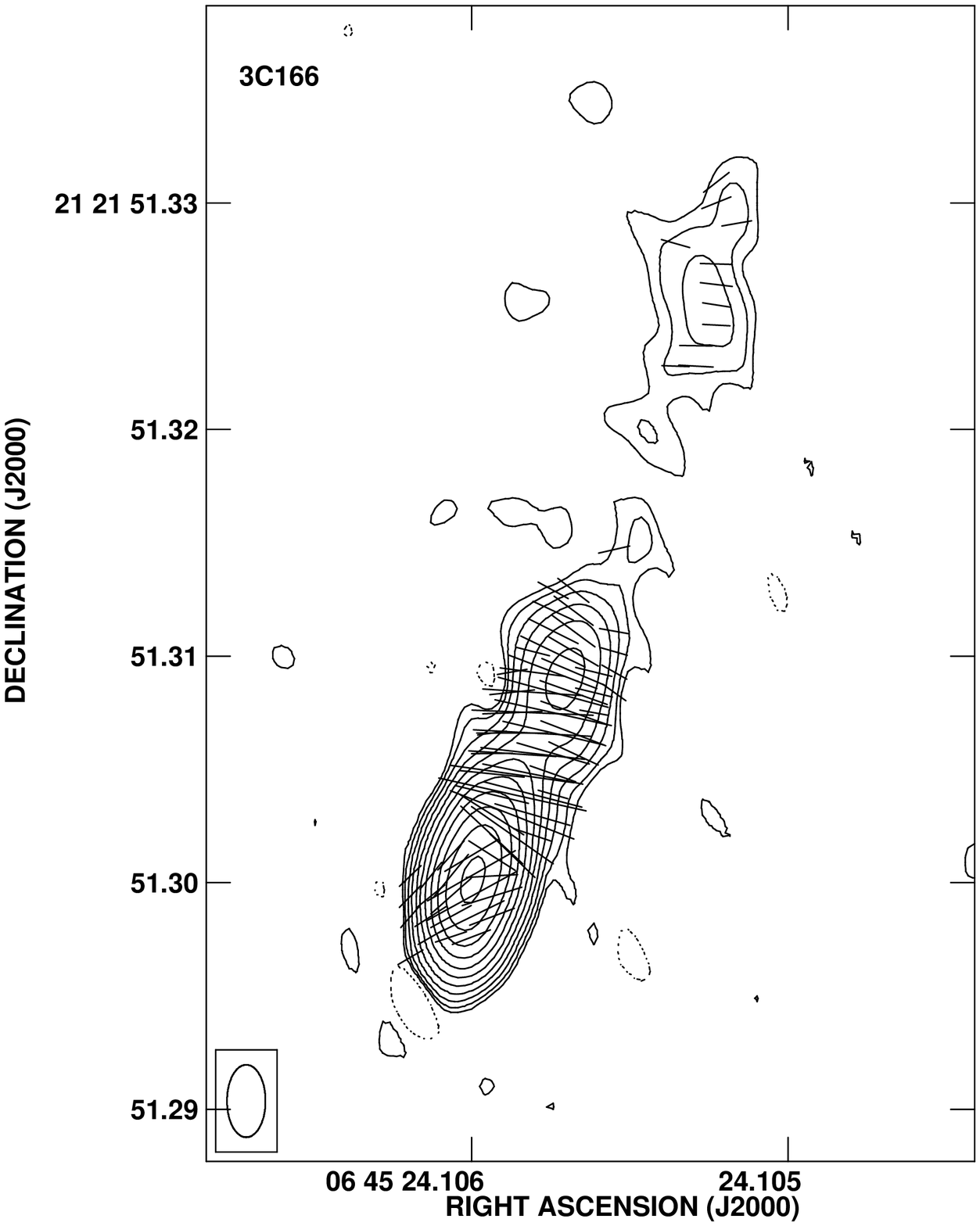}
\caption{Contour plot of 3C\,166 with electric polarization vectors at 
4882 MHz 
overlaid.  Contours start at 0.2 mJy/beam and increase by factors of
2.  The peak in the map is 240 mJy/beam.  
A polarization vector length of 1 mas corresponds to 0.167 mJy of 
polarized flux density.  No correction for Faraday rotation has been
made to the polarization angles.  The restoring beam 
has dimensions 3.2 $\times$ 1.7 mas in p.a. 0\deg.
\label{fig2}}
\end{figure}
\clearpage


\begin{figure}
\vspace{19cm}
\includegraphics{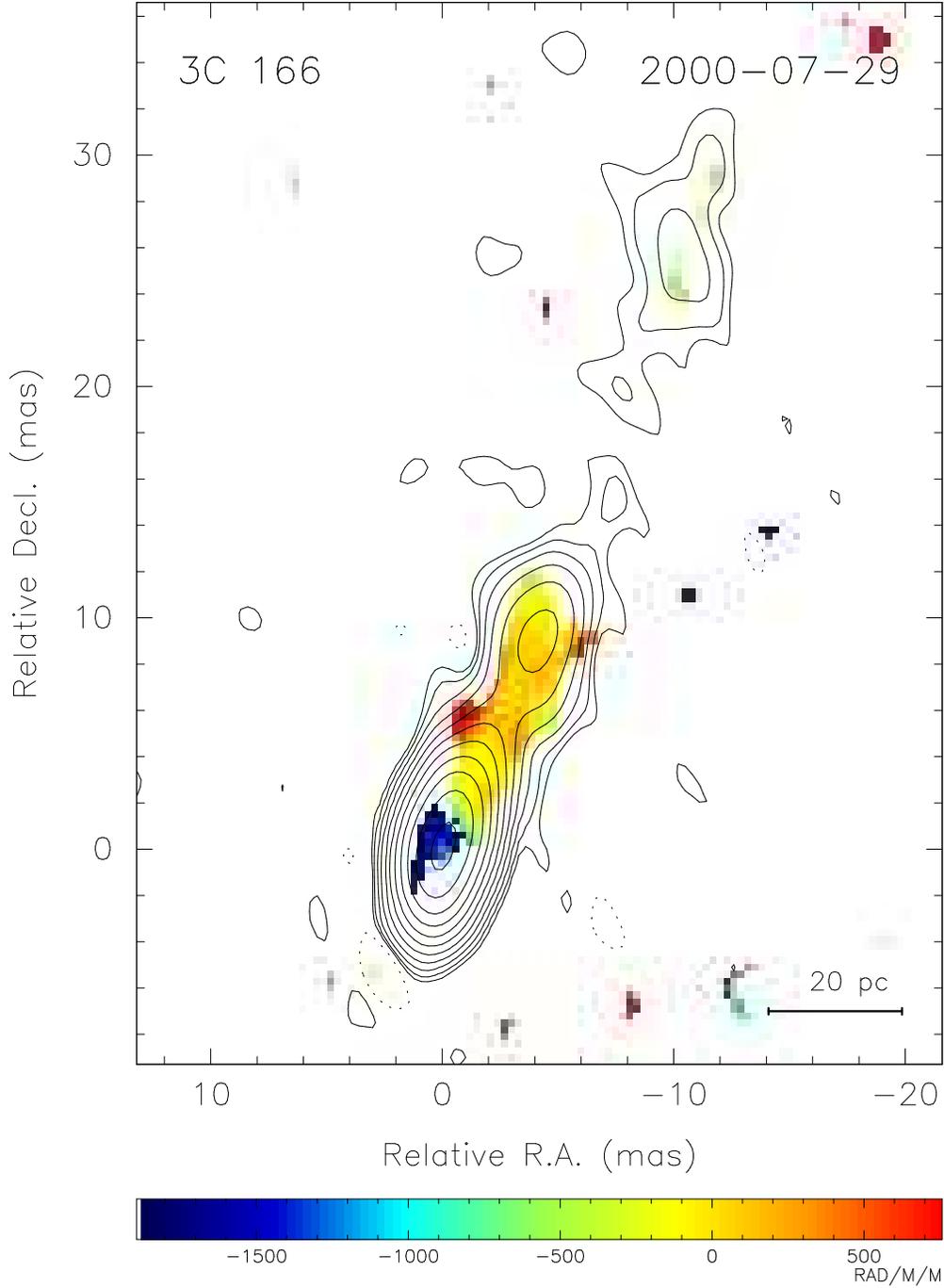}
\caption{Rotation measure image of 3C\,166.  Rotation Measures in 
the rest frame of the source are larger by a factor (1+z)$^2$ or 1.55. Contours are the same as
in Fig.~1.}
\label{fig3}
\end{figure}
\clearpage

\begin{figure}[t]
\vspace{19cm}
\includegraphics{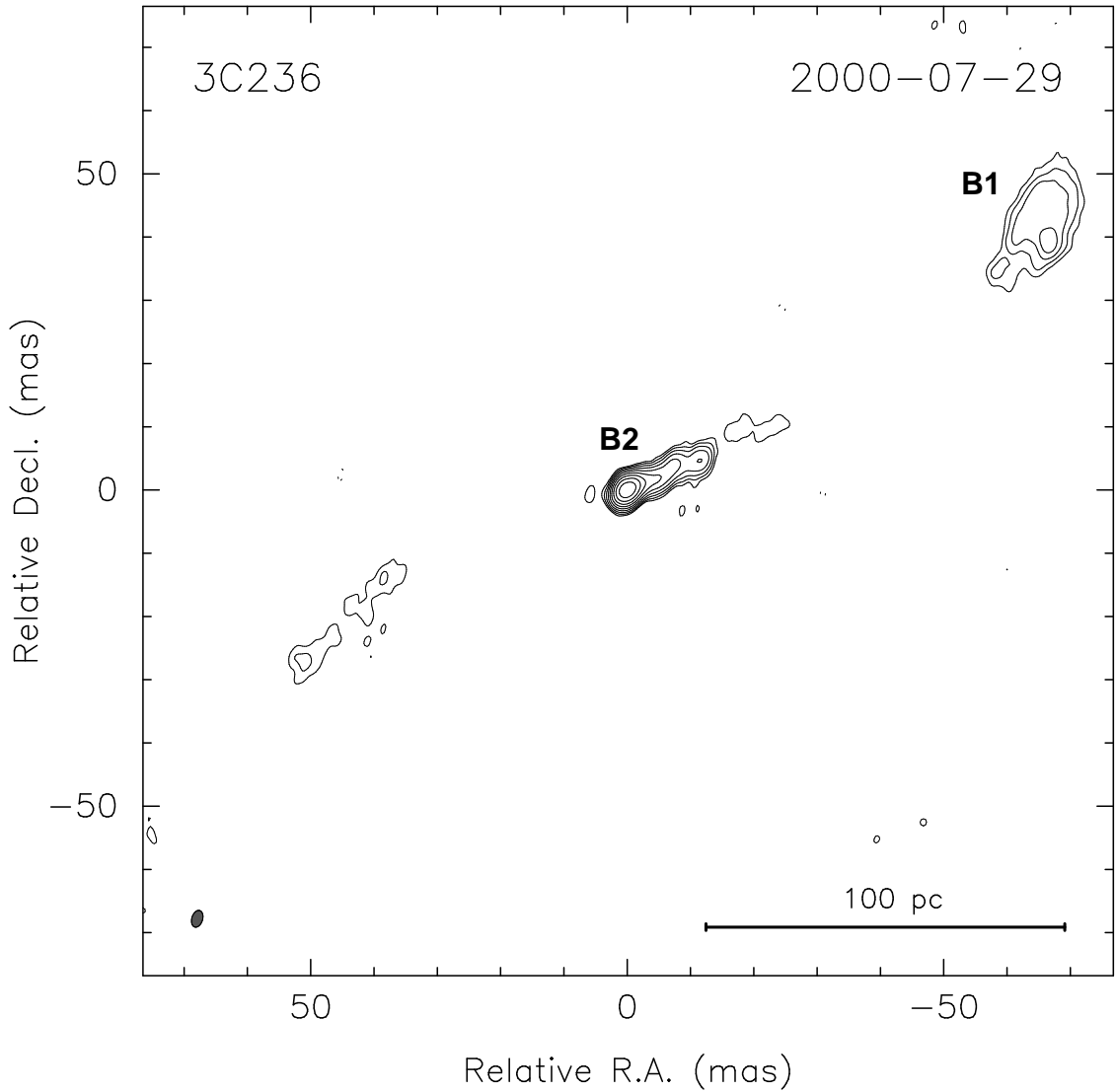}
\caption{Contour plot of the central 150 $\times$ 150 mas of 3C\,236.  Contours start at 0.3 mJy/beam and
increase by powers of 2. The restoring beam is drawn in the lower
left corner and has dimensions 2.7 $\times$ 1.7 mas in p.a. $-$17\deg.
The peak in the map is 136 mJy/beam.  }
\label{fig4}
\end{figure}
\clearpage

\begin{figure}[t]
\vspace{19cm}
\includegraphics{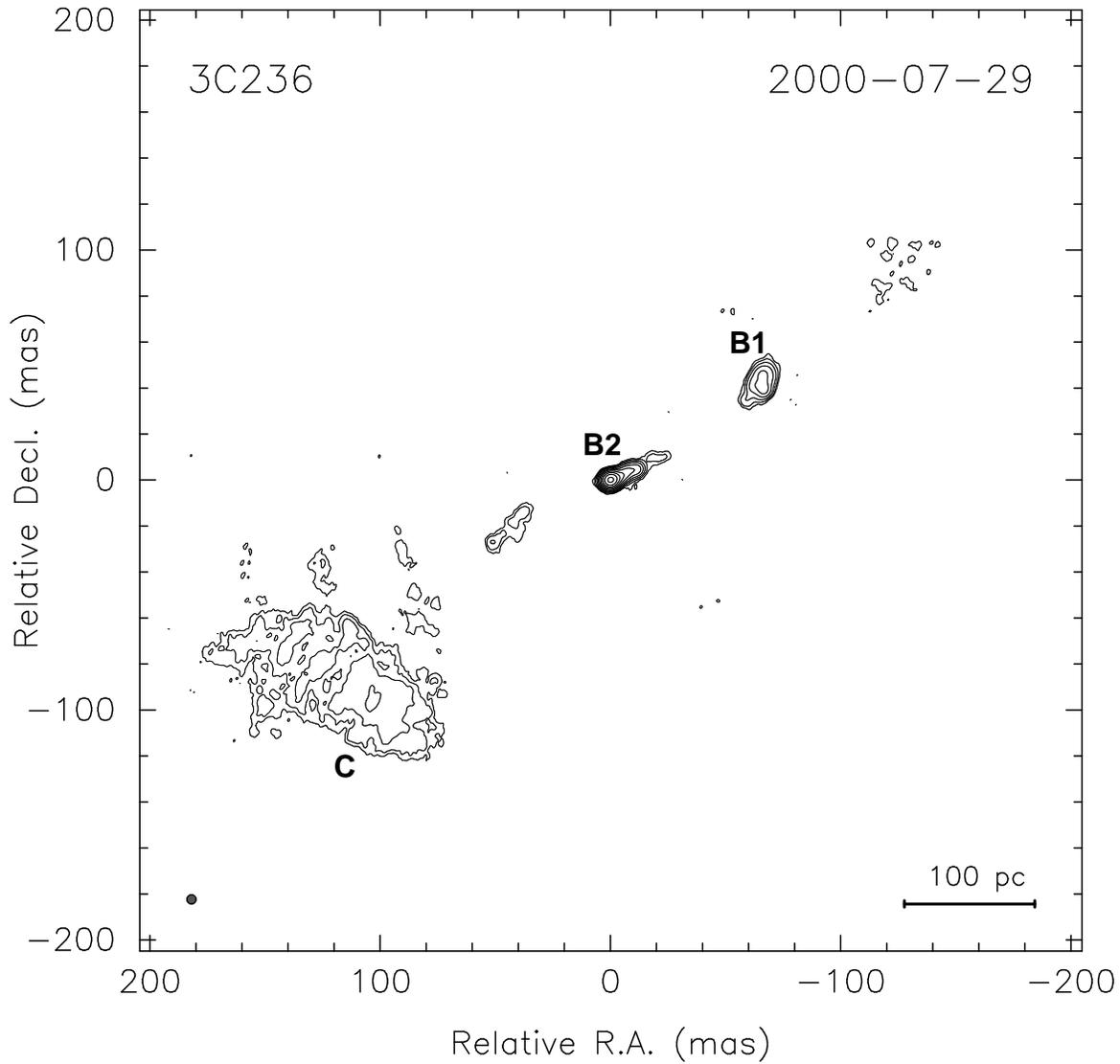}
\caption{Contour plot of the central 300 $\times$ 300 mas of 3C\,236.  A
circular restoring beam of 4 mas FWHM has been used.   Contours start at 0.3 mJy/beam and
increase by powers of 2.  The peak in the map is 190 mJy/beam.}
\label{fig5}
\end{figure}
\clearpage


\begin{figure}[t]
\vspace{19cm}
\includegraphics{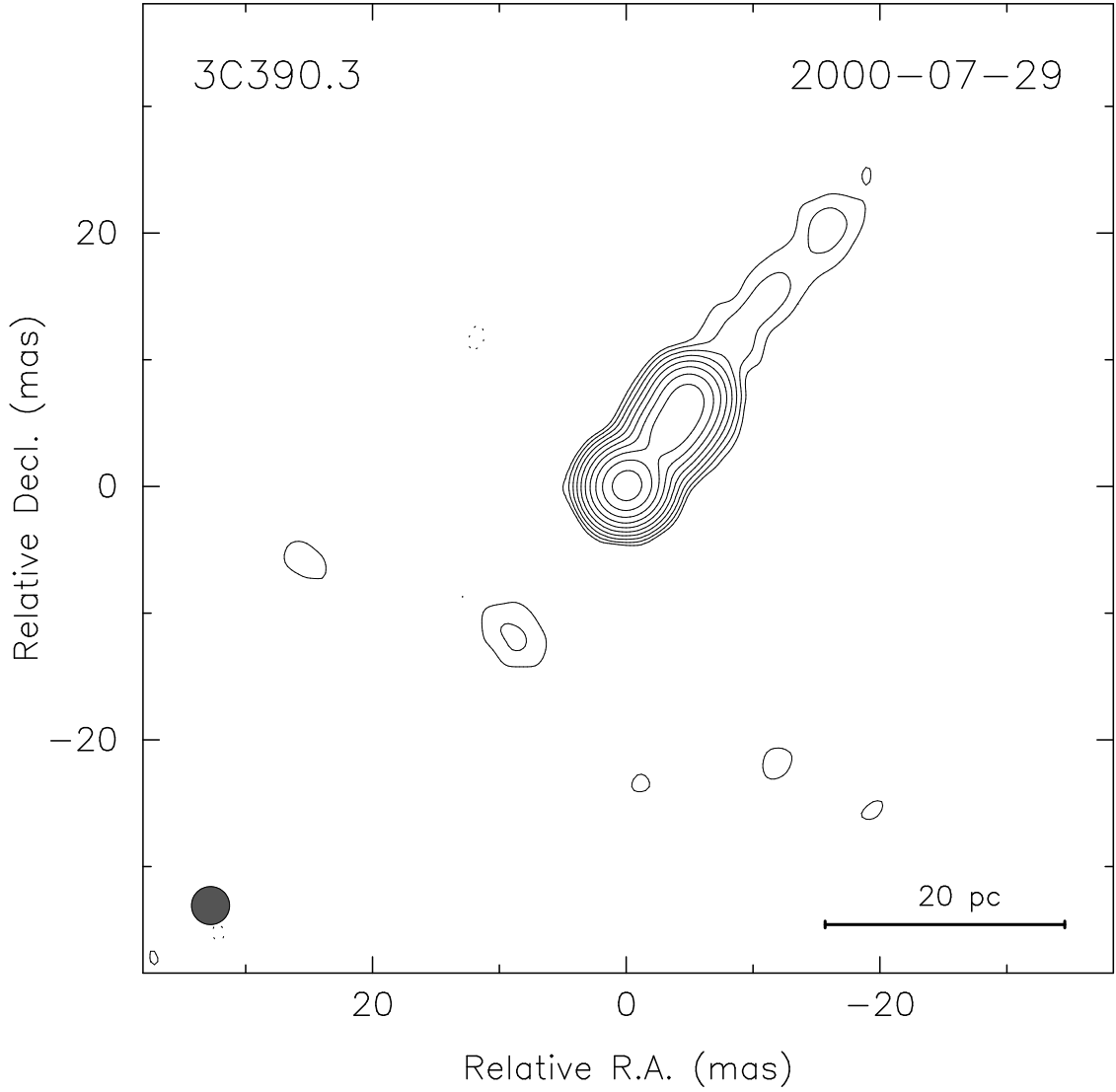}
\caption{Contour plot of 3C\,390.3.  Contours start at 0.25 mJy/beam and
increase by powers of 2.  A ($u,v$) taper of 70 M$\lambda$ has been
applied and a circular restoring beam of 3 mas used. The peak in 
the map is 190 mJy/beam. }
\label{fig7}
\end{figure}
\clearpage

\begin{center}
\tightenlines
\singlespace
TABLE 1 \\
\smallskip
VLBA O{\sc bservational} P{\sc arameters}
\smallskip
\begin{tabular}{l r r r r r r r r}
\hline
\hline
Source & Frequency & BW & Scans & Time \\
(1) & (2) & (3) & (4) & (5) \\
\hline
\noalign{\vskip2pt}
targets \\
3C\,166    & 4.517, 4.547, 4.867, 4.897 & 4 & 10 & 400 \\ 
3C\,236    & 4.517, 4.547, 4.867, 4.897 & 4 & 10 & 400 \\ 
3C\,390.3  & 4.517, 4.547, 4.867, 4.897 & 4 & 11 & 440 \\ 
calibrators \\
3C\,279    & 4.517, 4.547, 4.867, 4.897 & 4 & 3 & 9 \\ 
1308+326 & 4.517, 4.547, 4.867, 4.897 & 4 & 4 & 14 \\ 
OQ208    & 4.517, 4.547, 4.867, 4.897 & 4 & 10 & 30 \\ 
1749+096 & 4.517, 4.547, 4.867, 4.897 & 4 & 3 & 9 \\ 
1928+738 & 4.517, 4.547, 4.867, 4.897 & 4 & 11 & 35 \\ 
\hline
\end{tabular}
\end{center}
\begin{center}
{\sc Notes to Table 1}
\end{center}
Col.(1).---Source name.
Col.(2).---Observing frequency in GHz.
Col.(3).---Total spanned bandwidth in MHz.
Col.(4).---Number of scans (each 4 -- 40 minutes duration).
Col.(5).---Total integration time on source in minutes.
\bigskip
\clearpage

\begin{center}
TABLE 2 \\
\smallskip
P{\sc olarization} A{\sc ngle} C{\sc alibration}
\smallskip
\begin{tabular}{l r r r r r r r r}
\hline
\hline
Source & $\nu$ & $S_{\rm VLA}$ & $P_{\rm VLA}$ & $\chi_{\rm VLA}$ & $S_{\rm
VLBA}$ & $P_{\rm VLBA}$ & $\chi_{\rm VLBA}$ \\
(1) & (2) & (3) & (4) & (5) & (6) & (7) & (8) \\
\hline
\noalign{\vskip2pt}
1308+326 & 4.9 & 1.768 & 58 & $-$13 & 1.633 & 54 & $-$14 \\
1749+096 & 4.9 & 2.330 & 122 & $-$6 & 2.063 & 103 & $-$3 \\
\hline
\end{tabular}
\end{center}
\smallskip
\begin{center}
{\sc Notes to Table 2}
\end{center}
Col.(1).---Source name.
Col.(2).---Observing band in GHz.
Col.(3).---Integrated VLA flux density in Jy.
Col.(4).---Integrated VLA polarized flux density in mJy.
Col.(5).---VLA polarization angle (E-vector) in degrees.
Col.(6).---Integrated VLBA flux density in Jy.
Col.(7).---Integrated VLBA polarized flux density in mJy.
Col.(8).---VLBA polarization angle (E-vector) in degrees.
\bigskip

\bigskip

\begin{center}
TABLE 3 \\
\smallskip
S{\sc ource} P{\sc roperties} \\
\smallskip
\begin{tabular}{l c c c}
\hline
\hline
Property & 3C\,166 & 3C\,236 & 3C\,390.3 \\
\hline
\noalign{\vskip2pt}
core RA (J2000) & 06$^h$45$^m$24\rlap{$^s$}{.\,}106  & 10$^h$06$^m$01\rlap{$^s$}{.\,}756 & 18$^h$42$^m$09\rlap{$^s$}{.\,}035 \\ 
\phantom{Core}Dec. (J2000) &  21\arcdeg 21\arcmin 51\rlap{\arcsec}{.\,}300 & 34\arcdeg 54\arcmin 10\rlap{\arcsec}{.\,}460 &
 79\arcdeg 46\arcmin 17\rlap{\arcsec}{.\,}200 \\ 
redshift & 0.2449 & 0.0988 & 0.0561 \\
largest angular size & 50\arcsec & 42\arcmin & 3.7\arcmin  \\
largest physical size & 170 kpc & 4380 kpc &  235 kpc \\
flux density (4.9 GHz)$^a$ & 1229 mJy & 1672 mJy & 4380 mJy \\
power (4.9 GHz) & 18.0 $\times$ 10$^{25}$ W Hz$^{-1}$ & 3.9 $\times$ 10$^{25}$ W Hz$^{-1}$ & 3.1 $\times$ 10$^{25}$ W Hz$^{-1}$ \\
inner $\beta\cos\theta$ (VLBI jet) & 0.67 & 0.48 & 0.76 \\
inner $\theta_{\rm max}$ (VLBI jet) & 48 & 61 & 41 \\
\hline
\end{tabular}
\end{center}
$^a$ Total flux density from the 87 GB catalog of Gregory \& Condon (1991),
or in the case of 3C\,390.3 from K\"uhr \etal (1981).

\clearpage
\end{document}